# Microwave and spin transfer torque driven coherent control in ferromagnets

Marina Brik*, Nirel Bernstein, Amir Capua

Department of Applied Physics, The Hebrew University of Jerusalem, Jerusalem 91904, Israel

*e-mail: marina.brik@mail.huji.ac.il

**Abstract:**

**Coherent control is a method used to manipulate the state of matter using oscillatory electromagnetic radiation which relies on the non-adiabatic interaction. It is commonly applied in quantum processing applications. This technique is interesting in the context of ferromagnetic materials because of the ability to combine it with spintronics for the purpose of fundamental spin transport research, low-power information processing, and potentially future quantum bit (Qubit) applications. In this work we address the theoretical grounds of coherent manipulation in practical ferromagnetic systems. We study electromagnetic radiation driven interaction that is enhanced in the presence of spin polarized currents and map the conditions that allow coherent manipulation for which Rabi oscillations take place. The role of the magnetic anisotropy field is shown to act as an additional oscillatory driving field. We discuss the Gilbert losses in the context of effective coherence decay rates and show that it is possible to control these rates by application of a static spin current. The case of coherent manipulation using oscillatory spin currents that is free of radiation is discussed as well. Our work paves the way towards spin current amplification as well as radiation-free coherent control schemes that may potentially lead to novel Qubits that are robust and scalable.**

# I. INTRODUCTION

Coherent control is a method for controlling dynamical processes using electromagnetic radiation that translates a dynamical system from one state to another. At its basis stands the non-adiabatic dynamical regime. Studies of spin dynamics in magnetic media have been mainly carried out under either of the two regimes: the adiabatic regime for which a harmonic stimulus drives the system in its steady state (e.g. Refs. [1-3]), or under the free induction decay regime for which the response to an impulse is examined (e.g. Refs. [4-6]). These regimes have played an increasingly important role in understanding spin transport processes in atomically engineered solid state devices and key fundamental phenomena have been explored e.g. spin angular momentum losses [7-9], the spin Hall effect (SHE) [10-14], the anomalous Hall effect (AHE) [15], motion of magnetic domains [16-19], the spin transfer torques (STT) [20-22], and more.

A third dynamical regime that has so far received little attention in the context of ferromagnetic (FM) systems is the non-adiabatic regime. In this regime the system is driven by a microwave signal and the response is studied before steady state conditions prevail. Namely, the oscillatory driving field and the magnetization state exchange energy until the system decays to its steady precessional mode. When the energy is exchanged in a periodic manner Rabi oscillations that are characterized by the Rabi frequency take place and are the basis for coherent control and manipulation.

Because of the large gyromagnetic ratio of the electron, exploration of the non-adiabatic regime in ferromagnetic systems often requires fast electronics and/or synchronization circuitry capable of operating in the GHz range. Hence, experimental studies of the non-adiabatic regime in magnetic solid-state systems is usually more cumbersome. In the work by Karenowska (Ref. [23]), a spatial non-equilibrium energy exchange was demonstrated between counter propagating spin waves in yttrium iron garnets (YIG) using artificial magnonic crystals. In these experiments a periodic spatial modulation fulfilled the role of the oscillatory signal whereas the effect was recognized to be valuable for signal processing purposes. At the quantum limit, coherent control of single artificial magnetic spins was demonstrated using a scanning tunneling microscope (STM) [24]. In Ref. [24] magnetic Ti atoms were excited using microwaves to induce Rabi oscillations while initialization of the atoms was achieved by passing a DC spin current through the atom. This study was carried out in the time domain and the magnetization state was readout magnetoresistively.

Recently, we have demonstrated a hybrid time-frequency domain method to excite the non-adiabatic regime in magnetic media which we can describe as the pump-probe optically sensed ferromagnetic resonance (PP-OFMR) [25]. In this method, Rabi oscillations were excited in a few Å thick film of a CoFeB ferromagnet and in the presence of RF radiation following a perturbation by an intense ultrashort demagnetizing optical pulse. These experiments revealed a frequency chirp which was controllable by the static magnetic field and that the microwave field induced coherence in the inhomogeneously (IH) broadened spin ensemble. Moreover, the experiments showed that according to Gilbert's damping theory the intrinsic relaxation times were tunable by proper choice of the external magnetic field and when taken long enough they eventually initiated a resonant spin mode-locking of the system.

In the present work we provide the theoretical grounds for the non-adiabatic regime in a FM that is driven by microwave electromagnetic (EM) radiation and spin currents. We combine the formalism of the adiabatic interaction in FMs and approaches that are commonly used to describe two-state quantum systems. We compare between the representation of the spin-lattice and transverse spin polarization decay times, $T_1$ and $T_2^*$, used to describe quantum coherent phenomena and the Gilbert damping constant and IH broadening arising from Gilbert's damping theory. The influence of the injection of DC spin current on the effective coherence and decay times and the role of the magnetic anisotropy is studied. In addition, we present the case of using the AC spin current to drive the non-adiabatic dynamics instead of EM microwaves. We discuss the limitations of using AC STT as a driving force, and clarify its distinct nature compared to the ordinary RF EM field case.

Our work is presented as follows: we start by introducing the general conditions for observing Rabi oscillations in a magnetic system that is driven by a RF EM field. Next, we include the anisotropy fields and examine the case of the film having a perpendicular magnetic anisotropy (PMA) which is relevant for practical applications. We then add to our model a DC SHE. Specifically, we look into the influence of the injection of spin current on the overdamped and critically damped interactions. Finally, the non-adiabatic interaction is studied in the presence of a driving oscillatory STT generated by the SHE.

## II. MODEL AND RESULTS

### A. Model framework: Rabi oscillations in FMs

Our analysis is carried out under the framework of the macrospin approximation. To that end we start with the Landau-Lifshitz-Gilbert (LLG) equation for the magnetization, $\vec{M}'$, in the presence of the effective field, $\vec{H}'_{eff}$, in the lab frame of reference (indicated with a prime):

$$\frac{d\vec{M}'}{dt} = -\gamma(\vec{M}' \times \vec{H}'_{eff}) + \frac{\alpha}{M_s}\left(\vec{M}' \times \frac{d\vec{M}'}{dt}\right) \tag{1}$$

in which $M_s$ is the magnetization saturation, $\alpha$ is the Gilbert damping parameter, and $\gamma$ is the gyromagnetic ratio. In spherical coordinates the LLG equation converts to:

$$\begin{aligned}\dot{\theta}' &= \gamma H'_\varphi \\ \sin\theta' \cdot \dot{\varphi}' &= -\gamma H'_\theta\end{aligned} \tag{2}$$

where $H'_\theta$ and $H'_\varphi$ are the polar and azimuthal components of the effective field, respectively.

Following linearization we express the solution of Eq. (2) in a frame of reference rotating about the $\hat{z}'$ axis at the driving angular frequency, $\omega$, by $\theta(t) = \theta_0 + \Delta\theta(t)$ and $\varphi(t) = \varphi_0 + \Delta\varphi(t)$ with $(\Delta\theta(t), \Delta\varphi(t))$ being small deviations from equilibrium $(\theta_0, \varphi_0)$. $\Delta\theta(t)$ and $\Delta\varphi(t)$ are then expressed by their phasors $\Delta\theta = \Delta\theta_0 \cdot exp(-i\Omega t)$ and $\Delta\varphi = \Delta\varphi_0 \cdot exp(-i\Omega t)$, with $\Delta\theta_0$ and $\Delta\varphi_0$ being constants of the problem that are determined by the initial conditions. The complex frequency, $\Omega$, consists of the generalized Rabi frequency, $\Omega_R^G$, and the decay rate, $\Gamma$, according to $\Omega = -i\Gamma + \Omega_R^G$. Finally, $\Omega_R^G$ is given by $\Omega_R^G = \sqrt{\Omega_\sigma^2 - \Gamma^2}$ where $\Omega_\sigma$ and $\Gamma$ are obtained by satisfying the secular equation $\Omega^2 + i2\Omega\Gamma - \Omega_\sigma^2 = 0$ in the usual manner. Rabi oscillations are generally observable when the decay time is longer than the Rabi cycle, namely, when $\Gamma < \Omega_R^G$ and the response becomes underdamped.

### B. Fundamental interaction: EM driven dynamics

#### 1. Rabi frequency and linewidth

We first examine the microwave magnetic field driven interaction that will also serve as a reference case. The external magnetic field of magnitude $H_0$ is chosen in the $\hat{z}'$ direction while the oscillatory driving field of amplitude $h_{RF}$ is applied along the $\hat{x}'$ axis. In the rotating frame under the rotating wave approximation Eq. (2) becomes:

$$\begin{aligned}\dot{\theta} &= -\frac{1}{2}\gamma h_{RF} \cdot \sin\varphi - \alpha \cdot \sin\theta \cdot (\dot{\varphi} + \omega) \\ \sin\theta \cdot \dot{\varphi} &= \gamma\left(H_0 - \frac{\omega}{\gamma}\right) \cdot \sin\theta - \frac{1}{2}\gamma h_{RF} \cdot \cos\theta \cdot \cos\varphi + \alpha \cdot \dot{\theta}.\end{aligned} \tag{3}$$

$(\theta_0, \varphi_0)$ can be inferred from the equilibrium conditions $\dot{\theta} = \dot{\varphi} = 0$, while the time-dependent part of Eq. (3) gives:

$$\Delta\dot{\theta} = -\frac{1}{2}\gamma h_{RF} \cdot \cos\varphi_0 \cdot \Delta\varphi - \alpha \cdot \sin\theta_0 \cdot \Delta\dot{\varphi} - \alpha\omega \cdot \cos\theta_0 \cdot \Delta\theta$$

$$\sin\theta_0 \cdot \Delta\dot{\varphi} = \gamma\left(H_0 - \frac{\omega}{\gamma}\right) \cdot \cos\theta_0 \cdot \Delta\theta + \frac{1}{2}\gamma h_{RF} \cdot \cos\theta_0 \cdot \sin\varphi_0 \cdot \Delta\varphi \qquad (4)$$

$$+\frac{1}{2}\gamma h_{RF} \cdot \sin\theta_0 \cdot \cos\varphi_0 \cdot \Delta\theta + \alpha \cdot \Delta\dot{\theta}.$$

The set of Eq. (4) describe the conventional problem of a two-level system [26] with the difference that the spin angular momentum losses are incorporated through Gilbert's damping theory [27]. The Gilbert damping in the LLG equation has a rigorous physical origin. It originates from a Rayleigh friction process that is included to model losses such as those mediated by the spin-orbit and exchange interactions. Hence, the energy dissipation rate in our model is inherently dependent on numerous parameters of the problem with the most critical of them being the frequency of the precessional motion and consequently the external magnetic field [25]. In contrast, in the Bloch formalism the losses are incorporated through $T_1$ and $T_2^*$ and are generally independent of the effective field of the problem. The IH broadening that arises from variations in local anisotropy fields can be added in our model by taking variations in the effective bias field to the first order [28]. Figure 1 highlights the differences between the two models.

To calculate the Rabi flopping frequency, $\Omega_\sigma$ and $\Gamma$ can be determined from Eq. (5):

$$\Omega_\sigma = \gamma\left\{\frac{1}{(\alpha^2+1)}\left(\left(\frac{\alpha\omega}{\gamma}\right)^2 \cdot \cos^2\theta_0 + \left(H_0 - \frac{\omega}{\gamma}\right)^2 + \left(\frac{1}{2}h_{RF}\right)^2 \cdot \cos^2\varphi_0\right)\right\}^{\frac{1}{2}}$$

$$\Gamma = \gamma\frac{\alpha}{2(\alpha^2+1)}\left(\frac{2\omega}{\gamma} \cdot \cos\theta_0 + \left(H_0 - \frac{\omega}{\gamma}\right) \cdot \cos\theta_0 + \frac{1}{2}h_{RF} \cdot \sin\theta_0 \cdot \cos\varphi_0 + \frac{1}{2}h_{RF} \cdot \frac{\cos\varphi_0}{\sin\theta_0}\right). \qquad (5)$$

On resonance $\left(H_0 - \frac{\omega}{\gamma}\right) = 0$ and the solutions for $(\theta_0, \varphi_0)$ require $\theta_0 = +90°$ or $\varphi_0 = -90°$. The solution $\theta_0 = +90$ and $\varphi_0 = -90°$ corresponds to $\alpha\omega = \frac{1}{2}\gamma h_{RF}$, which indicates the transition from the overdamped to underdamped dynamics.

In the underdamped regime in which Rabi oscillations are observable, $\alpha\omega < \frac{1}{2}\gamma h_{RF}$, $\theta_0 = 90°$, and $\varphi_0 = -\arcsin(2\alpha\omega/\gamma h_{RF})$, resulting in:

$$\Omega_R^G = \gamma\sqrt{\left(\frac{1}{2}h_{RF}\right)^2 \cdot \cos^2(\varphi_0) \cdot \left(\frac{1}{(\alpha^2+1)} - \frac{\alpha^2}{(\alpha^2+1)^2}\right)}$$

$$\Gamma = \frac{\alpha}{2(\alpha^2+1)} \cdot \gamma h_{RF} \cdot \cos(\varphi_0) \qquad (6)$$

and when $\alpha\omega > \frac{1}{2}\gamma h_{RF}$, $\varphi_0 = -90°$ and $\theta_0 = \arcsin(\gamma h_{RF}/2\alpha\omega)$ and the response is overdamped with:

$$\Omega_R^G = \gamma \sqrt{\left(\frac{\alpha\omega}{\gamma}\right)^2 \cdot \cos^2(\theta_0) \cdot \left(\frac{1}{(\alpha^2+1)} - \frac{1}{(\alpha^2+1)^2}\right)} \tag{7}$$

$$\Gamma = \frac{\alpha\omega}{(\alpha^2+1)} \cdot \cos(\theta_0).$$

Figure 2(a) illustrates $\Omega_R^G$ and $\Gamma$ for $\alpha = 0.01$ and 10 GHz on resonance as a function of the normalized field $\frac{\gamma h_{RF}}{2\alpha\omega}$. The data resembles closely the dependence of the resonance frequency of a FM on the applied field when the external magnetic field is applied perpendicularly to the easy axis. In this case the quantity $\frac{1}{2}h_{RF}$ fulfills the same role as the static external field, $\frac{\alpha\omega}{\gamma}$ plays the role of the effective anisotropy field, and the easy axis is the rotation axis, $\hat{z}$. This effective anisotropy field arises from the projection of the Gilbert damping torque into the rotating frame and is hence dependent on both $\alpha$ and $\omega$, and similarly to an actual anisotropy field the torque that arises from it depends on the angle between the magnetization and the easy axis. For $\alpha\omega > \frac{1}{2}\gamma h_{RF}$, $\Gamma$ rapidly increases with decreasing $h_{RF}$ and becomes more than two orders of magnitude greater than $\Omega_R^G$ as shown in the inset so that Rabi oscillations are not obtained. When $\alpha\omega < \frac{1}{2}\gamma h_{RF}$ this behavior abruptly changes and $\Omega_R^G$ becomes much greater than $\Gamma$ with increasing $h_{RF}$ giving rise to Rabi nutations. Fig. 2(b) illustrates the difference $\Omega_R^G - \Gamma$ as a function of the detuning, $\left(H_0 - \frac{\omega}{\gamma}\right)$, and the normalized field from which the oscillatory nature can be determined. Starting from $\left(H_0 - \frac{\omega}{\gamma}\right) = \sim 30$ Oe the behavior is always oscillatory irrespective of $h_{RF}$. Figure 2(c) shows a typical temporal response for various $H_0$ values for which the response is overdamped at resonance and away from resonance becomes oscillatory.

## *2. Interpretation of the Gilbert damping torque in the rotating frame*

In the rotating frame of reference, the damping torque can be interpreted in a comprehensive manner providing further insight to the non-adiabatic interaction. In cartesian coordinates Eq. (1) transforms to:

$$\frac{d\vec{M}}{dt} = -\gamma \vec{M} \times \left(\left(\vec{H}_0 - \frac{\vec{\omega}}{\gamma}\right) + \vec{h}_{RF}\right) + \frac{\alpha}{M_s}\left(\vec{M} \times \frac{\delta \vec{M}}{\delta t}\right) - \frac{\alpha}{M_s}\vec{M} \times \left(\vec{M} \times \vec{\omega}\right) \tag{8}$$

where $\vec{\omega}$ is the vector $(0,0,\omega)$ and $\vec{h}_{RF}$ is the RF field. The first term on the right-hand side of Eq. (8) is the effective field $\left(\vec{H}_0 - \frac{\vec{\omega}}{\gamma}\right) + \vec{h}_{RF}$ which $\vec{M}$ primarily precesses about. The second term on the right-hand side of Eq. (8) is identical to the Gilbert damping term in the LLG equation and is responsible for the decay of the magnetic field towards the effective field. The third term, $-\frac{\alpha}{M_s}\vec{M} \times (\vec{M} \times \vec{\omega})$, does not appear in the LLG equation. It behaves as a non-conserving torque that has the form of the anti-damping STT term of the Landau-Lifshitz-Gilbert-Slonczewski (LLGS) equation. This term gives rise to the effective field $\frac{\alpha}{M_s}\left(\vec{M} \times \frac{\vec{\omega}}{\gamma}\right)$ and scales with $\alpha\omega$. In steady state, $\frac{\delta \vec{M}}{\delta t} = 0$, it balances the primary field $\left(\vec{H}_0 - \frac{\vec{\omega}}{\gamma}\right) + \vec{h}_{RF}$ and causes the system to decay towards a new steady state different than the one dictated solely by the primary field. Hence, this torque can be used as an additional control in a coherent manipulation scheme. Before steady state is reached its contribution to $\Omega_\sigma$ and hence also to $\Omega_R^G$ is readily seen in Eq. (5) where it appears as an additional term in the Euclidean norm of the fields consisting of $\vec{h}_{RF}$, $\left(\vec{H}_0 - \frac{\vec{\omega}}{\gamma}\right)$, and $\alpha\vec{\omega}/\gamma$ that eventually determine the Rabi frequency. This STT-like torque can be enhanced by increasing the driving frequency which is analogous to increasing the spin current in the LLGS equation.

*3. Large angle non-adiabatic interaction*

The analytical model addresses small deviations from equilibrium. At large angles of precession, the non-adiabatic response become nonlinear giving rise to the generation of higher harmonics. We examine this nonlinearity numerically [29-31]. A typical representative temporal response in the lab frame of reference is presented in Fig. 3 at 10 GHz and $h_{RF}$ of 90 Oe on resonance (following the experiments of [25]). In this example the magnetization was initialized to the $\hat{z}'$ direction and traversed the full swing towards the $-\hat{z}'$ direction. Fig. 3(a) shows the response for $\alpha$ of 0.01. The figure depicts $M_z'$ which is proportional to the energy of the system and $M_y'$. A nonlinear response consisting of higher harmonics is readily seen as well as an asymmetric behavior as $M_z'$ evolves, namely, as energy is absorbed or emitted. The 'down' transition is slower than the 'up' transition for which $\vec{M}'$ aligns with $\vec{H}_0'$. This is also readily seen on the $M_y'$ component which stretches or compresses in time depending on the 'up' or 'down' transition. At later times, as the response further decays and $\vec{M}'$ precesses at small angles near the steady state, the asymmetry vanishes and a harmonic response is revealed (Fig. 3(b)). This behavior is highly dependent on the Gilbert damping as shown in Fig. 3(c). When $\alpha$ is reduced to a value of 0.001 the nonlinearity vanishes and $M_z'$ oscillates at a single frequency according to the analytical model. The effect of the Gilbert damping on the nonlinear nature of the response is understood by examining the acting torques in a geometrical representation (Fig. 3(d)). The torque arising from the applied magnetic fields, namely RF and DC fields, $\frac{d\vec{M}'}{dt}\bigg|_{app.}$, can be decomposed into two components: a tangential component, $\frac{d\vec{M}'}{dt}\big|_{\parallel}$, responsible for the primary longitudinal precessional motion, and a transverse component, $\frac{d\vec{M}'}{dt}\big|_{\perp}$, responsible for the 'downward'/'upward' transition of $\vec{M}'$. When $\vec{M}'$ shifts towards $-\hat{z}'$, the transverse component, $\frac{d\vec{M}'}{dt}\big|_{\perp}$, is balanced by the Gilbert damping torque and the transition occurs at a slower rate. Likewise, when $\vec{M}'$ shifts towards $\hat{z}'$ the transverse component, $\frac{d\vec{M}'}{dt}\big|_{\perp}$, is enhanced by the Gilbert damping torque. Therefore, the Gilbert torque is responsible for the asymmetry in the 'upward'/'downward' transition rates. Finally, as $\vec{M}'$ further decays towards steady precessional state the applied torque is primarily tangential and the transverse torque component $\frac{d\vec{M}'}{dt}\big|_{\perp}$ is negligible resulting in a harmonic response. This behavior is more prominent the greater $\alpha$ is (Fig. 3(a), 3(c)). Hence, changes in $\alpha$ can be readily seen on the nonlinear non-adiabatic response thereby providing an additional way to investigate the loss mechanisms.

## 4. Inclusion of magnetic anisotropy

Practical magnetic systems exhibit magnetic anisotropy fields such as demagnetization and/or crystalline anisotropies. We focus on the case of a sample having perpendicular magnetic anisotropy (PMA) such as the geometry studied in [25] which is usually more important for technological purposes and allows high density and lower crosstalk between devices in practical applications.

The modeled geometry is illustrated in Fig. 4. The easy axis of magnetization is set along $\hat{y}'$. The effective anisotropy is $H_{Keff} = \frac{2K_u}{M_s} - 4\pi M_s$, where $K_u$ is the crystalline anisotropy constant. The analysis was carried out under the condition $H_0 > H_{Keff}$ for which the precession takes place around the $\hat{z}'$ axis in the lab frame. Under these conditions the problem has a closed form analytical solution. Hence, Eq. (4) become:

$$\begin{aligned}
\Delta\dot{\theta} &= -\frac{1}{2}\gamma h_{RF} \cdot \cos\varphi_0 \cdot \Delta\varphi - \alpha \cdot \sin\theta_0 \cdot \Delta\dot{\varphi} - \alpha\omega \cdot \cos\theta_0 \cdot \Delta\theta \\
\sin\theta_0 \cdot \Delta\dot{\varphi} &= -\frac{\gamma H_{Keff}}{2} \cdot \cos 2\theta_0 \cdot \Delta\theta + \gamma\left(H_0 - \frac{\omega}{\gamma}\right) \cdot \cos\theta_0 \cdot \Delta\theta \\
&\quad + \frac{1}{2}\gamma h_{RF} \cdot \cos\theta_0 \cdot \sin\varphi_0 \cdot \Delta\varphi + \frac{1}{2}\gamma h_{RF} \cdot \sin\theta_0 \cdot \cos\varphi_0 \cdot \Delta\theta \\
&\quad + \alpha \cdot \Delta\dot{\theta}
\end{aligned} \quad (9)$$

while $(\theta_0, \varphi_0)$ is found as before.

The solutions for $\Omega_\sigma$ and $\Gamma$ are:

$$\Omega_\sigma = \gamma \left\{ \frac{1}{(\alpha^2+1)} \left( \left(\frac{\alpha\omega}{\gamma}\right)^2 \cdot \cos^2\theta_0 - \frac{1}{2}h_{RF} \cdot \frac{1}{2}H_{Keff} \cdot \frac{\cos 2\theta_0 \cdot \cos\varphi_0}{\sin\theta_0} + \frac{1}{2}h_{RF} \cdot \left(H_0 - \frac{\omega}{\gamma}\right) \cdot \right. \right.$$

$$\left. \left. \cdot \frac{\cos\theta_0 \cdot \cos\varphi_0}{\sin\theta_0} + \left(\frac{1}{2}h_{RF}\right)^2 \cdot \cos^2\varphi_0 \right) \right\}^{\frac{1}{2}}$$

$$\Gamma = \gamma \frac{\alpha}{2(\alpha^2+1)} \left( \frac{2\omega}{\gamma} \cdot \cos\theta_0 - \frac{H_{Keff}}{2} \cdot \cos 2\theta_0 + \left(H_0 - \frac{\omega}{\gamma}\right) \cdot \cos\theta_0 + \frac{1}{2}h_{RF} \cdot \sin\theta_0 \cdot \right.$$

$$\left. \cdot \cos\varphi_0 + \frac{1}{2}h_{RF} \cdot \frac{\cos\varphi_0}{\sin\theta_0} \right).$$

(10)

For small $\alpha$, small $h_{RF}$, and quasi resonance conditions $\left(H_0 - \frac{\omega}{\gamma}\right) = 0$, $\varphi_0 = -180°$ and the condition for $\theta_0$ is $\sin\theta_0 = h_{RF}/H_{Keff}$, for small $h_{RF}$. Substituting the above conditions into $\Omega_\sigma$ term in Eq. (9), we get:

$$\Omega_R^G = \gamma \sqrt{\left(\frac{H_{Keff}}{2}\right)^2 - \left(\frac{h_{RF}}{2}\right)^2}. \qquad (11)$$

Equation (11) has the familiar form of resonance frequency for the PMA case where the external field is applied perpendicular to the easy axis. Thus, in the rotating frame $h_{RF}$ takes the role of DC field applied perpendicularly to the easy axis. In comparison with the PMA case [32], $\frac{1}{2}H_{Keff}$ appears as an effective anisotropy field in the rotating frame. The dependence of $\Omega_R^G$ on $h_{RF}$ in Eq. (11) is fundamentally different from its general dependence in standard two-level systems. Here $\Omega_R^G$ decreases with increasing $h_{RF}$ whereas in conventional two-level systems $\Omega_R^G$ increases with the driving field amplitude. In conventional two-level systems the driving field amplitude determines the rate at which the occupation probabilities evolve, therefore $\Omega_R^G$ generally increases with $h_{RF}$. In contrast, in Eq. (11) $\Omega_R^G$ decreases with $h_{RF}$ and is due to the role played by the anisotropy field which effectively acts as an additional oscillatory driving field. Therefore, the conventional dependence breaks until $h_{RF}$ reaches the limit of $\frac{1}{2}H_{Keff}$. The numerical model shows this behavior as well. The temporal responses presented in Fig. 4 shows that $\Omega_R^G$ decreases as $h_{RF}$ increases up to 60 Oe ($\frac{1}{2}H_{Keff}$). Above 60 Oe $h_{RF}$ overcomes the anisotropy field and $\Omega_R^G$ increases with $h_{RF}$ in the expected manner.

### C. Interaction in the presence of DC spin current

From a technological point of view, a static STT may play an important role in the non-adiabatic interaction in magnetic systems because it can be used to actively tune the decay rates according to the LLGS equation [21,22]. Specifically, STT can be used to extend the coherence time of the system making the FM system a versatile platform for coherent control schemes. In the model derived hereon we consider the anti-damping like STT and assume that the spin current is generated by the SHE in a heavy metal-ferromagnet bilayer [33]. Hence, a DC charge current, $J_c$, is applied along the $\hat{x}'$ direction and generates a spin current density $J_s\hat{y}'$ with spin angular momentum aligning in the $\hat{z}'$ direction. The presence of spin current introduces the torque $\frac{\gamma H_{SHE,DC}}{M_s}\left(\vec{M}' \times (\vec{M}' \times \hat{s}')\right)$ into Eq. (1). Here $\hat{s}'$ is a unit vector in the direction of the injected spin angular momentum and $H_{SHE,DC}$ is the SHE parameter defined by $H_{SHE,DC} = \frac{\hbar \theta_{SH} J_c}{2eM_s t_{FM}}$ where $\hbar$ is the reduced Planck constant, $e$ is the electron charge, $\theta_{SH}$ is the spin Hall angle (SHA), and $t_{FM}$ is the thickness of the FM layer into which the spin current is injected. With these substitutions Eq. (4) becomes:

$$\Delta\dot{\theta} = -\frac{1}{2}\gamma h_{RF} \cdot \cos\varphi_0 \cdot \Delta\varphi - \alpha \cdot \sin\theta_0 \cdot \Delta\dot{\varphi} - (\alpha\omega - \gamma H_{SHE,DC}) \cdot \cos\theta_0 \cdot \Delta\theta$$
$$\sin\theta_0 \cdot \Delta\dot{\varphi} = \gamma\left(H_0 - \frac{\omega}{\gamma}\right) \cdot \cos\theta_0 \cdot \Delta\theta + \frac{1}{2}\gamma h_{RF} \cdot \cos\theta_0 \cdot \sin\varphi_0 \cdot \Delta\varphi \quad (12)$$
$$+ \frac{1}{2}\gamma h_{RF} \cdot \sin\theta_0 \cdot \cos\varphi_0 \cdot \Delta\theta + \alpha \cdot \Delta\dot{\theta}$$

resulting in $\Omega_\sigma$ and $\Delta\Omega$:

$$\Omega_\sigma = \gamma\left\{\frac{1}{(\alpha^2+1)}\left(\left(H_{SHE,DC} - \frac{\alpha\omega}{\gamma}\right)^2 \cdot \cos^2\theta_0 + \left(H_0 - \frac{\omega}{\gamma}\right)^2 + \left(\frac{1}{2}h_{RF}\right)^2 \cdot \cos^2\varphi_0\right)\right\}^{\frac{1}{2}}$$
$$\Gamma = \gamma\frac{1}{2(\alpha^2+1)}\left(2\left(\frac{\alpha\omega}{\gamma} - H_{SHE,DC}\right) \cdot \cos\theta_0 + \alpha\left(H_0 - \frac{\omega}{\gamma}\right) \cdot \cos\theta_0 + \alpha\frac{1}{2}h_{RF} \cdot \sin\theta_0 \cdot \quad (13)\right.$$
$$\left. \cdot \cos\varphi_0 + \alpha\frac{1}{2}h_{RF} \cdot \frac{\cos\varphi_0}{\sin\theta_0}\right).$$

Equation (13) shows that the spin current compensates the Gilbert damping term according to the difference $\frac{\alpha\omega}{\gamma} - H_{SHE,DC}$. Eq. (13) reveals that for the critical case of $H_{SHE,DC} = \frac{\alpha\omega}{\gamma}$ which may be achieved in realistic systems having SHA of 0.15 e.g. Pt, W [11,12] the response is always underdamped, namely, $\Gamma < \Omega_R^G$, and Rabi oscillations appear irrespective of the magnitude of $h_{RF}$. Obviously, these oscillations still decay as $\Gamma \neq 0$. To examine the influence of the injected spin current on the existence of Rabi oscillations and the damping rate we explored the interaction under resonance conditions. Figure 5 illustrates the temporal responses from which the transition between the overdamped and underdamped regimes for various charge current levels is seen. The figure shows $M_Z$ as a function of $h_{RF}$ with $\alpha = 0.01$, $M_s = 300$ emu/cm$^3$, $\theta_{SH} = 0.15$ and $t_{FM} = 11.5$ Å corresponding to Ref. [33]. Figure 5(a) presents the case with no spin current. It is seen that the transition between the overdamped and underdamped responses occurs at $h_{RF} = 71\ Oe$ above which Rabi oscillations take place. When the DC current is increased the threshold reduces up to the point of $H_{SHE,DC} = \frac{\alpha\omega}{\gamma}$ (Fig. 5(c)) in which the oscillations are obtained for any value of $h_{RF}$. As the DC current is further increased, the threshold in $h_{RF}$ increases again (Fig. 5(d)-(f)). This behavior stems from the fact that when $\alpha < \frac{\gamma H_{SHE,DC}}{\omega}$ the term $2\gamma \left( \frac{\alpha\omega}{\gamma} - H_{SHE,DC} \right) \cdot \cos\theta_0$ in Eq. (13) adds a positive contribution to the damping rate. Thus, at high DC currents the Rabi oscillations eventually become overdamped. Most importantly, Fig. 5 shows that when Rabi oscillations take place the coherence times can be tuned by the DC spin current. This is seen from the varying decay rates as marked by the guiding lines in Fig. 5(a)-(c). It is seen that as $J_c$ increases to the critical value ($2.5 \cdot 10^6 \, A/cm^2$ in our case) the coherence times extend.

### D. AC STT driven non-adiabatic dynamics

Inclusion of AC charge current creates an AC STT which serves as an alternative driving force. A driving force of this kind is advantageous over the RF driven case for scalability purposes since it does not require a radiating micro-antenna but only physical contact to the device. However, the AC STT driving force has a different nature compared to the ordinary Zeeman oscillatory magnetic field. We include the AC STT by replacing $\vec{h}'_{RF}$ in Eq. (1) with an AC charge current density $\vec{J}'_c = J_0 \cdot \cos(\omega t) \cdot \hat{y}'$. The AC charge current is converted by the SHE to an AC spin current density $J_s \hat{z}'$ having $\hat{s}'$ in the $\hat{x}'$ direction which introduce a STT term of $\frac{\gamma h_{SHE,AC}}{M_s} \left( \vec{M}' \times (\vec{M}' \times \hat{s}') \right)$ in Eq. (1). $h_{SHE,AC}$ is the SHE parameter as in section II.C, that refers now to an AC current. The $\vec{J}'_c$ direction was chosen such that $\hat{s}'$ is orthogonal to the static magnetization equilibrium vector in the lab frame $\vec{M}'_0$. Following linearization, the AC STT in the lab frame equals approximately $\frac{\gamma h_{SHE,AC}}{M_s} \left( \vec{M}'_0 \times (\vec{M}'_0 \times \hat{s}') \right)$. Thus, if $\vec{M}'_0$ and $\hat{s}'$ are collinear, the AC STT vanishes. The time-dependent equations become:

$$\begin{aligned}
\Delta\dot{\theta} &= \frac{1}{2}\gamma h_{SHE,AC} \cdot \cos\theta_0 \cdot \sin\varphi_0 \cdot \Delta\varphi + \frac{1}{2}\gamma h_{SHE,AC} \cdot \sin\theta_0 \cdot \cos\varphi_0 \cdot \Delta\theta \\
&\quad - \alpha \cdot \sin\theta_0 \cdot \Delta\dot{\varphi} - \alpha\omega \cdot \cos\theta_0 \cdot \Delta\theta \\
\sin\theta_0 \cdot \Delta\dot{\varphi} &= \gamma\left(H_0 - \frac{\omega}{\gamma}\right) \cdot \cos\theta_0 \cdot \Delta\theta + \frac{1}{2}\gamma h_{SHE,AC} \cdot \cos\varphi_0 \cdot \Delta\varphi + \alpha \cdot \Delta\dot{\theta}
\end{aligned} \qquad (13)$$

and $\Omega_\sigma$ and $\Gamma$ for this case are given by:

$$\Omega_\sigma = \gamma \left\{ \frac{1}{(\alpha^2+1)} \left( \left(\frac{\alpha\omega}{\gamma}\right)^2 + \left(H_0 - \frac{\omega}{\gamma}\right)^2 \cdot \cos^2\theta_0 + \left(\frac{1}{2}h_{SHE,AC}\right)^2 \cdot \cos^2\varphi_0 \right) \right\}^{\frac{1}{2}} \qquad (15)$$

$$\Gamma = \gamma \frac{1}{2(\alpha^2+1)} \left( \frac{\alpha\omega}{\gamma} \cdot \cos\theta_0 + 2\alpha\left(H_0 - \frac{\omega}{\gamma}\right) \cdot \cos\theta_0 - \frac{1}{2} h_{SHE,AC} \cdot \sin\theta_0 \cdot \right.$$
$$\left. \cdot \cos\varphi_0 - \frac{1}{2} h_{SHE,AC} \cdot \frac{\cos\varphi_0}{\sin\theta_0} \right).$$

To understand the general behavior of the solution we assume very small α and resonance conditions. From Eq. (15) it is seen that $\Gamma$ takes a nonzero value even when $\alpha \to 0$, thus, the AC STT contributes to the decay term. From steady state we find $(\theta_0, \varphi_0) = (90°, 180°)$. Substituting $(\theta_0, \varphi_0)$ into Eq. (15) we get $\Omega_R^G = 0$. Hence, the response is overdamped (Fig. 6(a)) regardless of the $h_{SHE,AC}$ value even in the absence of damping in contrast to the $h_{RF}$ driven case. It can be further verified that the overdamped response persists in the vicinity of resonance as long as $\left|\left(H_0 - \frac{\omega}{\gamma}\right)\right| < \frac{1}{2} h_{SHE,AC}$. This distinctly different behavior of the AC STT compared to the RF magnetic field case can be understood by observing Eq. (8) in the rotating frame in which $\vec{h}_{RF}$ appears in the primary torque term $-\gamma \vec{M} \times \left(\left(\vec{H}_0 - \frac{\vec{\omega}}{\gamma}\right) + \vec{h}_{RF}\right)$. In contrast, when the system is driven solely by AC STT the primary torque vanishes on resonance leaving only the AC damping-like STT. For this reason, an additional DC STT cannot excite Rabi oscillations under resonance conditions, but only change the steady state. Away from resonance conditions and for $\left|\left(H_0 - \frac{\omega}{\gamma}\right)\right| > \frac{1}{2} h_{SHE,AC}$ Rabi oscillations are observable. When losses are included in addition, a DC STT applied in the geometry of section II.C affects the losses in the same manner as in the $\vec{h}_{RF}$ case where it extends the coherence time as long as $H_{SHE,DC} \leq \frac{\alpha\omega}{\gamma}$. When $H_{SHE,DC} > \frac{\alpha\omega}{\gamma}$, the coherence time decreases and for relatively high $H_{SHE,DC}$ values the oscillations are totally suppressed. In the current analysis, the field-like term of the AC STT was neglected since in many material systems it is much smaller than the damping-like term. However, when the field-like term is not negligible, it can excite Rabi oscillations even on resonance because its form is identical to the $\vec{h}_{RF}$ torque and hence appears in the primary precessional torque term of Eq. (8). Figure 6(b) shows the temporal responses of $M_z$ as a function of $H_0$. It is readily seen that despite the differences between the $\vec{h}_{RF}$ and the AC STT cases especially on resonance, the AC STT driven dynamics qualitatively behave in the same manner as the $\vec{h}_{RF}$ driven interaction (Fig. 2(c)).

Finally, when the magnetic anisotropy is included, it can be verified that for the same geometry introduced earlier (section II.B.4), Eq. (15) take the form:

$$\Omega_\sigma = \gamma \left\{ \frac{1}{(\alpha^2+1)} \left( \left(\frac{\alpha\omega}{\gamma}\right)^2 - \frac{1}{2} h_{SHE,AC} \cdot \left(H_0 - \frac{\omega}{\gamma}\right) \cdot \frac{\cos^2\theta_0 \cdot \sin\varphi_0}{\sin\theta_0} + \frac{1}{2} h_{SHE,AC} \cdot \frac{1}{2} H_{keff} \cdot \right. \right.$$

$$\left. \left. \cdot \frac{\cos 2\theta_0 \cdot \cos\theta_0 \cdot \sin\varphi_0}{\sin\theta_0} + \left(\frac{1}{2} h_{SHE,AC}\right)^2 \cdot \cos^2\varphi_0 \right) \right\}^{\frac{1}{2}} \quad (16)$$

$$\Gamma = \gamma \frac{1}{2(\alpha^2+1)} \left( \frac{\alpha\omega}{\gamma} \cdot \cos\theta_0 + \alpha \left(H_0 - \frac{\omega}{\gamma}\right) \cdot \cos\theta_0 - \alpha \cdot \frac{1}{2} H_{keff} \cdot \cos 2\theta_0 - \alpha \frac{1}{2} h_{SHE,AC} \cdot \right.$$

$$\left. \cdot \frac{\cos\theta_0 \cdot \sin\varphi_0}{\sin\theta_0} - \frac{1}{2} h_{SHE,AC} \cdot \sin\theta_0 \cdot \cos\varphi_0 - \frac{1}{2} h_{SHE,AC} \cdot \frac{\cos\varphi_0}{\sin\theta_0} \right).$$

For $\alpha \to 0$ and resonance conditions the steady state equations give $\varphi_0 = 90°$ and the condition for $\theta_0$ is $\sin 2\theta_0 = h_{SHE,AC}/0.5 H_{Keff}$, for $h_{SHE,AC} < H_{Keff}$. Inserting these conditions into Eq. (16), we get: $\Omega_R^G = \sqrt{\Omega_\sigma^2 - \Gamma^2} = \sqrt{\frac{1}{2}\gamma h_{SHE,AC} \cdot \frac{1}{2}\gamma H_{keff} \cdot \frac{\cos 2\theta_0 \cdot \cos\theta_0}{\sin\theta_0}}$. Thus, when the anisotropy fields are included, $\Omega_R^G \neq 0$ and Rabi oscillations take place on resonance (quasi-resonance).

### III. SUMMARY

In this work we examined the non-adiabatic interaction which is the basis for coherent control schemes in magnetic materials and relied on a hybrid two-level/adiabatic interaction in FMs formalism. We explored the ordinary non-adiabatic interaction driven by RF field and mapped the conditions for reaching the Rabi oscillations for which coherent control is made possible. We studied the energy transfer rates and showed that at large angles of precession and large $\alpha$ values the absorption and emission rates become highly non-symmetric. Furthermore, this nonlinear non-adiabatic response provided an additional way to investigate the loss mechanisms. We demonstrated that it is possible to control the effective coherence time by the injection of DC current and explored the non-adiabatic interaction in a system driven by an alternative driving source, namely, the AC STT, and concluded that there are no Rabi oscillations on resonance, as long as the AC STT field-like term is negligible. However, it is possible to get on resonance Rabi oscillations if the field-like term is non-negligible and this can motivate the search for magnetic materials that possess significant STT field-like term. Extensions of our work include complementing the existing experimental work to fully map the non-adiabatic regime in FM systems as well as to discuss a truly coherent control scheme that relays on the principles outlined here as well as novel coherent spin current amplification schemes (to be discussed in a follow-up paper). Further into the future, STT can be utilized as a versatile platform for coherent control schemes to be used in the manipulation of Qubits.

**Figure 1**

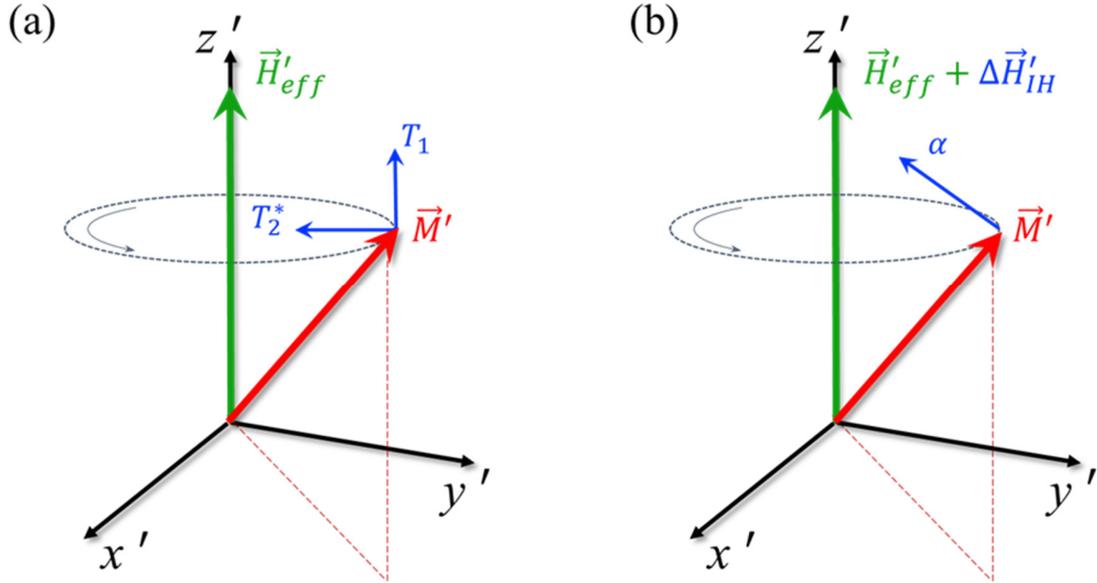

Figure 1. Geometrical representation of the damping and relaxation torques and the IH broadening of Bloch-Bloembergen and Gilbert pictures. (a) Bloch-Bloembergen picture. Blue arrows represent the $T_1$ and $T_2^*$ lattice and transverse relaxation torques, respectively. $\vec{H}'_{eff}$ is the effective magnetic field (b) Gilbert damping picture. Gilbert damping torque is indicated by the blue arrow and the IH broadening, $\Delta\vec{H}'_{IH}$, is included through variations in $\vec{H}'_{eff}$.

**Figure 2**

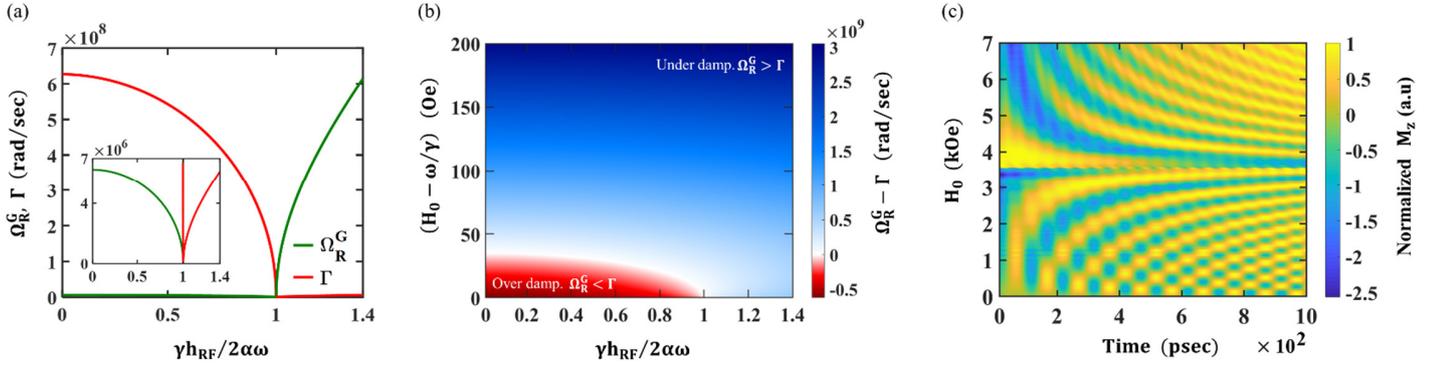

Fig 2. (a) $\Omega_R^G$, $\Gamma$ for zero detuning. Inset shows close-up of $\Omega_R^G$ at $\gamma h_{RF}/2\alpha\omega < 1$. (b) $\Omega_R^G$-$\Gamma$ as a function of the detuning. The red zone indicates the overdamped region, while the blue zone is the underdamped region. (c) Temporal responses as a function of $H_0$ calculated numerically for $h_{RF}$=20 Oe. The slight high frequency modulation observed as a background arises from the counter rotating terms that are neglected in the model. Results are presented for $\alpha = 0.01$ and 10 GHz.

**Figure 3**

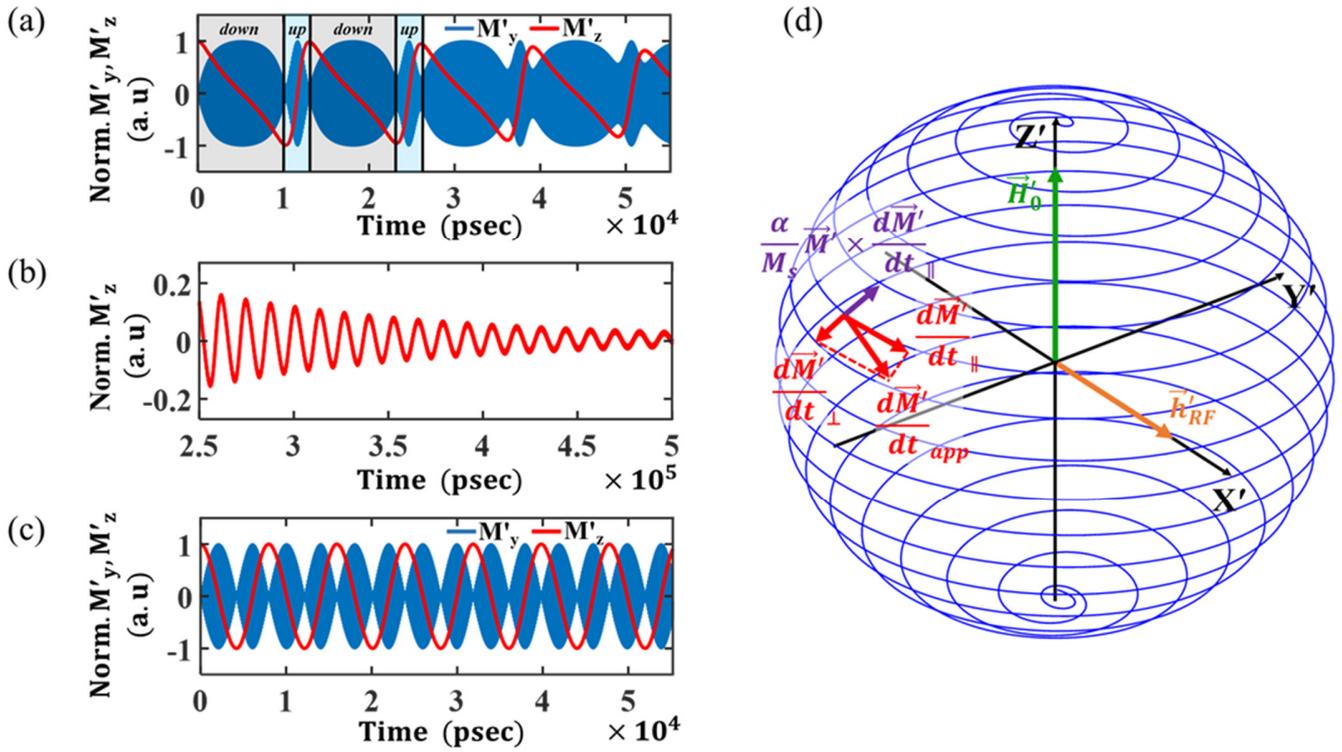

Fig 3. Large signal response calculated numerically. (a) Response calculated at 10 GHz and $\alpha = 0.01$. (b) Same simulation presented in (a) at later times. (c) Same conditions as in (a) but with $\alpha = 0.001$. (d) Geometrical representation of the torques acting on the magnetic moment in the lab frame.

**Figure 4**

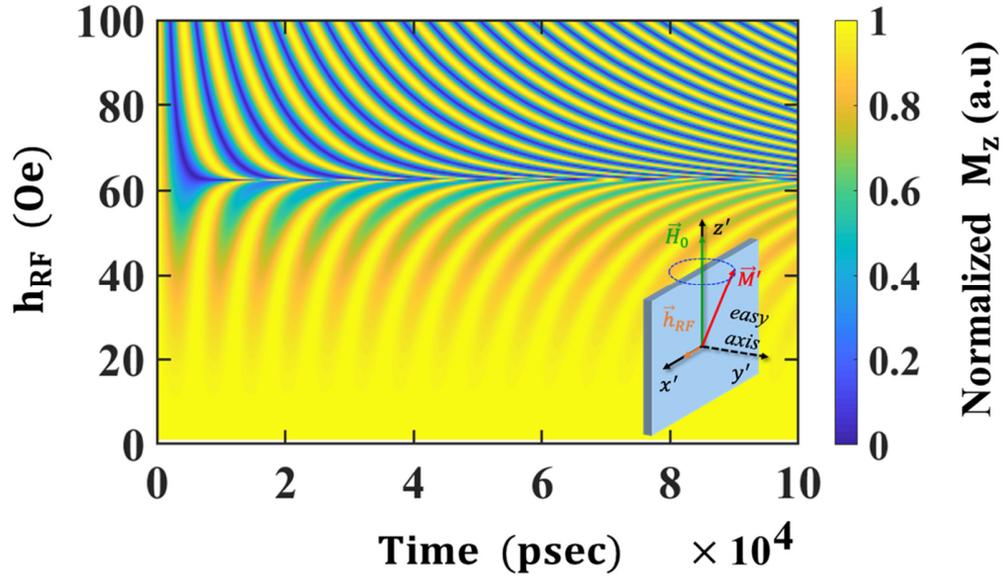

Fig 4. Temporal $M_z$ responses as a function of $h_{RF}$ calculated numerically for $H_0 = \frac{\omega}{\gamma}$ (quasi-resonance), effective anisotropy field of 120 Oe, and $\alpha = 10^{-4}$. Inset shows the PMA modeled sample geometry.

**Figure 5**

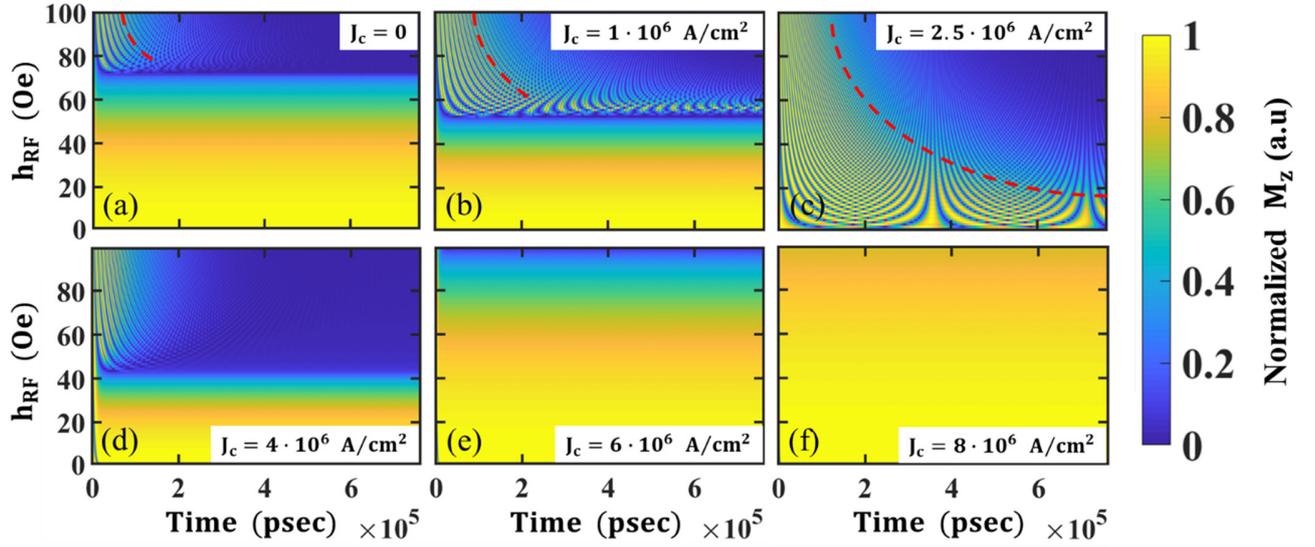

Fig 5. Temporal responses as a function of $h_{RF}$ calculated numerically for $J_C$ levels of $0\ A/cm^2$ to $8 \cdot 10^6 A/cm^2$ ((a) to (f)). The red dashed guiding lines indicate the varying decay rates.

**Figure 6**

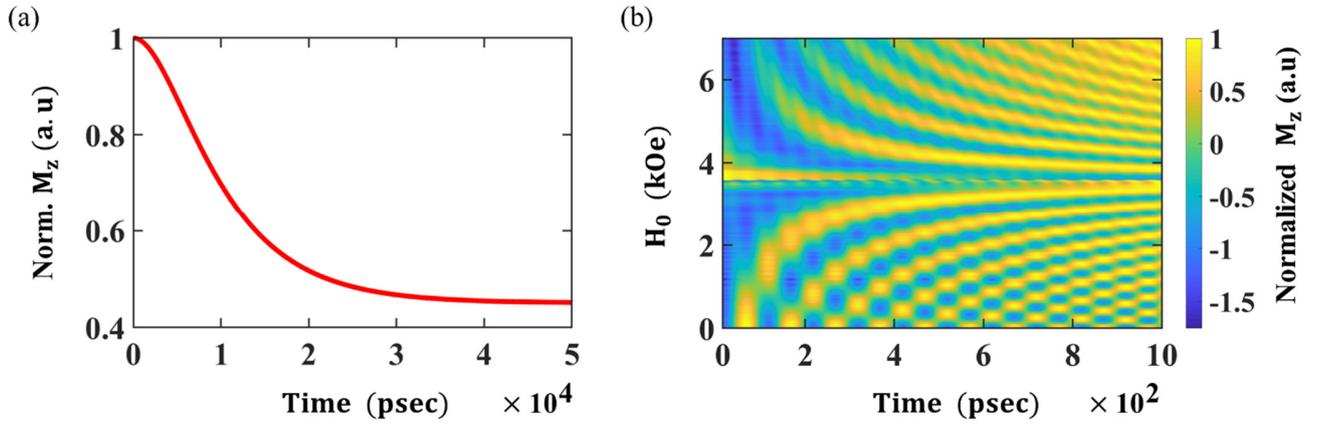

Figure 6. Temporal responses of a system driven by AC STT. (a) Response on resonance $\left(H_0 - \frac{\omega}{\gamma}\right) = 0$ for AC charge current density $J_C$ with amplitude of $1 \cdot 10^6 A/cm^2$ and $\alpha = 0.001$. (b) Temporal responses as a function of $H_0$ calculated numerically for the same $J_C$ value as in (a) and $\alpha = 0.005$.